\documentclass[aps,twocolumn,raggedbottom,prl,showpacs,nobalancelastpage,amssymb,superscriptaddress]{revtex4} 
\usepackage[english]{babel}
\usepackage{graphicx}
\usepackage{amssymb}
\usepackage{amsmath}
\usepackage{amscd}
\usepackage{eucal}
\usepackage{color}
\usepackage{bm}
\usepackage{upgreek}

\usepackage{epstopdf}

\def\beq{\begin{equation}}
\def\eeq{\end{equation}}
\def\beqa{\begin{eqnarray}}
\def\eeqa{\end{eqnarray}}
\def\bsub{\begin{subequations}}
\def\esub{\end{subequations}}

\def\rmd{\mathrm{d}}
\def\rme{\mathrm{e}}
\def\rmi{\mathrm{i}}
\def\Imag{\mathrm{Im}}
\def\Real{\mathrm{Re}}
\def\Tr{\mathrm{Tr}}

\newcommand{\tE}{\tau_{\rm E}}
\newcommand{\tD}{\tau_{\rm D}}

\begin{document}

\title{Semiclassical Gaps in the Density of States of Chaotic Andreev Billiards}
\author{Jack Kuipers}
\author{Daniel Waltner}
\author{Cyril Petitjean}
\affiliation{Institut f\"ur Theoretische Physik, Universit\"at Regensburg, D-93040 Regensburg, Germany}
\author{Gregory Berkolaiko}
\affiliation{Department of Mathematics, Texas A\&M University, College Station, TX 77843-3368, USA}
\author{Klaus Richter}
\affiliation{Institut f\"ur Theoretische Physik, Universit\"at Regensburg, D-93040 Regensburg, Germany}


\date{\today}

\begin{abstract}
The connection of a superconductor to a chaotic ballistic quantum dot leads to interesting phenomena, 
most notably the appearance of a hard gap in its excitation spectrum.  Here we treat such an 
Andreev billiard semiclassically where the density of states is expressed in terms of the classical 
trajectories of electrons (and holes) that leave and return to the superconductor. We show how 
classical orbit correlations lead to the formation of the hard gap, as predicted by random matrix theory in the limit of negligible Ehrenfest time $\tE$, and how the influence of a finite $\tE$ causes the gap to shrink.  Furthermore, for intermediate $\tE$ we predict a second gap below $E\!=\!\pi \hbar /2\tE$
which would presumably be the clearest signature yet of $\tE$-effects.
\end{abstract}


\maketitle

A superconductor (S) in contact with a normal conductor (N) considerably affects its
spectral density of quasiparticle excitations: due to Andreev reflection~\cite{And64} 
at the NS interface the density of states (DoS) is suppressed closely above the Fermi energy
$E_{\rm F}$. This proximity effect is also expected for an `Andreev billiard'~\cite{Kos95},
an impurity-free quantum dot attached to a superconductor~\cite{Mor97,Jak00}, and has attracted
considerable theoretical attention during the last decade (see [\onlinecite{Bee05}] for a review). 

An Andreev billiard has the interesting peculiarity that the suppression of its (mean)
DoS crucially depends on whether the dynamics of its classical counterpart is
integrable or chaotic: while the DoS vanishes linearly in energy for the
integrable case, the spectrum of a chaotic billiard is expected to exhibit 
a true gap above $E_{\rm F}$~\cite{Mel96}.   Based on random matrix theory (RMT) this gap was predicted to scale with the Thouless energy, $E_{\rm T} \!=\! \hbar /2\tD$, where $\tD$ is the 
average (classical) dwell time a particle stays in the billiard between successive Andreev 
reflections~\cite{Mel96}. On the contrary, semiclassics based on the so-called Bohr-Sommerfeld 
(BS) approximation yields only an exponential suppression of the DoS~\cite{Lod98,Sch99,Ihr01},
a discrepancy that has attracted much theoretical interest~\cite{Tar01,Ada02b,Sil03,Vav03, Mic09}.
Lodder and Nazarov~\cite{Lod98} pointed out that these seemingly contradictory predictions
are valid in different limits, governed by the ratio $\tau \!=\! \tE/\tD$. Here the
(quantum mechanical) Ehrenfest time $\tE \sim |\ln\hbar|$ separates the evolution of wave 
packets following essentially the classical dynamics from longer time scales dominated by 
wave interference.  In the universal regime, $\tau \!=\! 0$, the Thouless gap (from RMT) is clearly 
established~\cite{Mel96,Tar01}, while the BS approximation describes the classical 
limit $\tau \!\to\! \infty$.

Various approaches have been used to better understand the crossover from the
Thouless to the Ehrenfest regime of large $\tau$, where RMT loses its applicability~\cite{Tar01}.
These include effective RMT~\cite{Sil03}, predicting a gap size scaling with the Ehrenfest energy 
$E_{\rm E}\! =\! \hbar/2\tE$, as well as stochastic~\cite{Vav03} and perturbative~\cite{Ada02b} 
methods. Recently the gap at $\pi E_{\rm E}$ was confirmed for $\tau\!\gg\!1$ in a quasiclassical approach based on the Eilenberger equation \cite{Mic09}.

The purpose of this Letter is twofold. Firstly, using the scattering approach~\cite{Bee91}, we demonstrate that the DoS can be evaluated semiclassically for $\tE \! = \! 0$ by using an energy-dependent extension of the work~\cite{bhn08} on the moments of the transmission eigenvalues.  This semiclassically computed DoS yields a hard gap, in agreement with RMT.
Secondly we address the whole crossover regime of $\tau \! > \! 0$, by incorporating the $\tE$ dependence.
In the limit $\tau \! \gg \! 1$, the width of the gap approaches $\pi E_{\rm E}$, eventually recovering the BS prediction
for $\tau \! \to \! \infty$. More interestingly in the intermediate regime $\tau \! \ge \! 1$ we predict the appearance of a second `Ehrenfest' gap at $\pi E_{\rm E}$.

{\em Andreev billiard.}---In the scattering approach the superconductor is represented by a lead that
carries $M$ scattering channels, and the excitation spectrum can be entirely expressed in 
terms of the (electron) scattering matrix $S$~\cite{Bee91}. The average DoS reads~\cite{Ihr01} 
(when divided by twice the average density of the isolated billiard),
\beq \label{densityeqn}
d(\epsilon)=1+2\sum_{n=1}^{\infty}\frac{(-1)^{n}}{n}\Imag \frac{\partial C(\epsilon,n)}{\partial \epsilon},
\eeq
in terms of correlation functions of $n$ $S$-matrices,
\beq \label{Cetaneqn}
C(\epsilon,n) \!=\!\frac{1}{M}\Tr\left[S^{\dagger}\left(E_{\rm F} \!-\!
\frac{\epsilon\hbar}{2\tD}\right)S\left(E_{\rm F}
\!+\!\frac{\epsilon\hbar}{2\tD}\right)\right]^n\! , 
\eeq
at different energies. Here the energy difference $2E$ is expressed in units of the Thouless energy
and $\tD = T_{\mathrm{H}}/M$ with $T_{\mathrm{H}}$ the Heisenberg time, {\em i.e.}\ the time conjugate to the mean level spacing. For $\epsilon\!=\!0$, the $C(\epsilon,n)$ in Eq.~(\ref{Cetaneqn}), with transmission rather than scattering matrices, give the moments of the transmission eigenvalues which were calculated semiclassically (to leading order in $M^{-1}$) in \cite{bhn08} and our derivation is based on that work.

\begin{figure} [b]
\includegraphics[width=0.95\columnwidth]{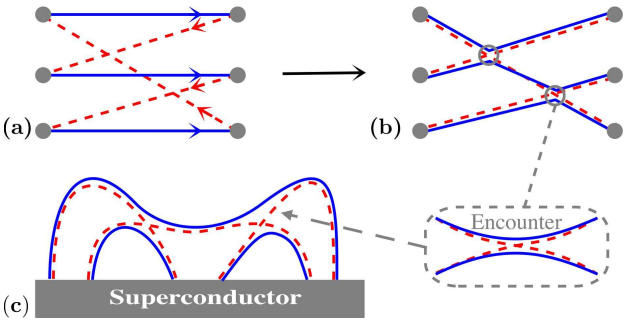}
\caption{\label{4trajpic} (a) Schematic picture of the trajectory structures for $n \! = \! 3$. 
The (blue) solid lines represent electrons which are retro-reflected as holes [dashed (red) lines]. 
(b) Collapsing the trajectories onto each other leads to encounters. 
(c) The end result, {\em i.e.}\ correlated Andreev reflected paths.}
\end{figure}

{\em Semiclassical evaluation in the universal re\-gime.}---To evaluate Eq.~(\ref{Cetaneqn}), we start with the semiclassical approximation to the  
scattering matrix elements connecting the channel $a$ to $b$, which are given by \cite{miller75}
\beq \label{scatmateqn}
S_{ba}(E_{\rm F}\pm E) \approx \frac{1}{\sqrt{T_{\mathrm{H}}}}\sum_{\zeta (a \to b)}
A_{\zeta}\rme^{\frac{\rmi}{\hbar}S_{\zeta}(E_{\rm F}\pm E)},
\eeq
in terms of the classical trajectories $\zeta$ connecting $a$ to $b$. Here $S_{\zeta}$ is the
action of $\zeta$, and $A_\zeta$ is its stability (including Maslov indices).  We
substitute Eq.~(\ref{scatmateqn}) into Eq.~(\ref{Cetaneqn}) and expand the action up to first
order in the energy yielding the duration $T_{\zeta}=\partial S_{\zeta}/\partial E_{\rm F}$.  The
correlators  are then given by a sum over $2n$ trajectories
\begin{align} \label{Csemieqn}
& C(\epsilon,n)  \approx \frac{1}{M {T_{\mathrm{H}}}^{n}}\prod_{j=1}^{n} \sum_{a_{j},b_{j}} 
 \sum_{\zeta_{j}(a_{j}\to b_{j})} \sum_{ \zeta_{j}' (b_{j}\to a_{j+1})} A_{\zeta_{j}}A_{\zeta_{j}'}^{*} \nonumber\\
& \times \exp{[(\rmi/\hbar)(S_{\zeta_{j}}\!-\!S_{\zeta_{j}'})]}
  \exp{[(\rmi\epsilon/2\tD) (T_{\zeta_{j}}+T_{\zeta_{j}'})]}, 
\end{align}
with $a_{n+1}=a_{1}$. The final trace of the product of matrices means that the trajectories complete a cycle, moving forward along the $\zeta_{j}$ and back along the $\zeta_{j}'$; for an example of this structure for $n\!=\!3$, see Fig.~\ref{4trajpic}(a).

In Eq.~(\ref{Csemieqn}) we add the actions of all the unprimed trajectories
and subtract the actions of the primed ones, so the resulting phase oscillates wildly,
unless the total action difference is of the order of $\hbar$.  One way to get small action
differences is to collapse all the trajectories onto each other [see
Fig.~\ref{4trajpic}(b)].  This leads to encounters where the electron trajectories
avoid crossing while the hole trajectories cross (or vice versa) to ensure that they each connect
the correct channels. In phase space, the trajectories of course do not cross, but just come close enough together to allow this reconnection.
Besides this direct collapse further possibilities arise from sliding 
the encounters together or into the leads (see \cite{bhn08}).

For each possibility we also need to know its semiclassical contribution. Following the treatment for
open systems of the first off-diagonal pair by \cite{rs02}, the
generalization to all orders \cite{heusleretal06} led to diagrammatic rules,
whereby each link ({\em i.e.}\ each trajectory stretch connecting channels or encounters)
essentially gives a factor of $[M(1-\rmi\epsilon)]^{-1}$, while each $l$-encounter (where $l$ electron trajectories come close together) contributes $-M(1-\rmi l\epsilon)$ as the encounter
stretches all remain inside the cavity or touch the lead together.  Summing the contributions, by extending the work of \cite{bhn08} to include energy differences \cite{Kui09details}, and using the diagrammatic rules above, we arrive at the intermediate generating function $g(\epsilon,r)$, which includes all possible diagrams apart from where the top encounter enters the lead, and is given implicitly by
\beq \label{ggenfunctioneqn}
g\left(1-\rmi\epsilon\right)-1=rg^2\left(g-1-\rmi\epsilon\right).
\eeq
Including the possibility where the top encounter can enter the lead, leads to the generating function 
\beq \label{defG}
G(\epsilon,r)=\sum_{n=1}^{\infty}r^{n-1}C(\epsilon,n)=\frac{g}{1-rg}
\eeq
of the correlation functions. By inverting Eq.~(\ref{defG}) we can see that $G$ is given
implicitly by the cubic equation
\beq \label{Gcubeqn}
r(r-1)^2G^3+r(3r+\rmi\epsilon-3)G^2 +(3r+\rmi\epsilon-1)G=-1.
\eeq
Expanding $G$ (or $g$) as a power series in $r$, we obtain the first couple of correlation functions (which can be checked by considering the semiclassical diagrams explicitly) as:
\beqa \label{Cepsexpansion}
C(\epsilon,1)&=&\frac{1}{\left(1-\rmi\epsilon\right)}, \quad C(\epsilon,2)=\frac{1-2\rmi\epsilon-2\epsilon^2}{\left(1-\rmi\epsilon\right)^4}, 
\eeqa

{\em Density of states in the universal regime.}---We can generate these correlation functions recursively to obtain a truncated version of the sum in 
Eq.~(\ref{densityeqn}).  However, we can go one step further and find the generating function of the terms that appear in the density of states
\beqa \label{Hgeneqn}
\nonumber H(\epsilon,r)&=& \frac{1}{\rmi r} \frac{\partial}{\partial \epsilon}\int G(\epsilon,r) \: \rmd r = \sum_{n=1}^{\infty}\frac{r^{n-1}}{\rmi n}\frac{\partial C(\epsilon,n)}{\partial \epsilon} ,\\
d(\epsilon)&=&1-2\Real \: H(\epsilon,-1) ,
\eeqa
using a computer aided search over cubic equations with low order polynomial coefficients to obtain
\beqa \label{Hcubeqn}
(\epsilon r)^2(1-r)H^3+\rmi\epsilon r\left[r(\rmi\epsilon-2)+2(1-\rmi\epsilon)\right]H^2 & & \nonumber \\
+\left[r(1-2\rmi\epsilon)-(1-\rmi\epsilon)^2\right]H+1=0 . &&
\eeqa
As $G$ and $H$ are solutions of algebraic generating functions, so too must be their derivatives, and we prove the result in Eq.~(\ref{Hcubeqn}) by differentiating $G$ with respect to $\epsilon$ and $rH$ with respect to $r$ and demonstrating that these derivatives satisfy the same algebraic equation.

\begin{figure}[b]
\centering
\includegraphics[width=0.95\columnwidth]{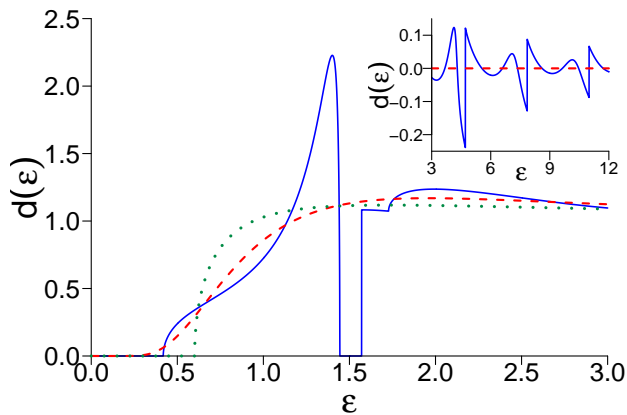}
\caption{Density of states for $\tau\!=\!\tE/\tD\! =\! 2$ (solid line), along with the BS (dashed) and 
RMT (dotted) limits, showing a second gap just below $\epsilon\!=\!\frac{\pi}{2}$. Inset: $\tE$-related
 $\pi$-periodic DoS oscillations at higher energy (after subtracting the BS curve).}
\label{Figtau2}
\end{figure}

Taking the solution whose expansion agrees with Eq.~(\ref{Cepsexpansion}), the DoS then follows from Eq.~(\ref{Hgeneqn}) as
\beq \label{densityRMT}
d(\epsilon)=\sqrt{3}/(6\epsilon)\left[Q_{+}(\epsilon)-Q_{-}(\epsilon)\right]
, \,\,  \epsilon>2[(\sqrt{5}-1)/2]^{\frac{5}{2}} ,
\eeq
where $Q_{\pm}(\epsilon)=\left[8-36\epsilon^2\pm3\epsilon\sqrt{3\epsilon^{4}+132\epsilon^{2}-48}\right]^{\frac{1}{3}}$. This is exactly the RMT result \cite{Mel96} (dotted line in Fig.~\ref{Figtau2}).

{\em Density of states in the Ehrenfest regime.}---The effect of non-zero Ehrenfest time $\tE$ on the first three correlation functions
$C(\epsilon, \tau, n)$ with $\tau = \tE / \tD$ has previously been calculated 
semiclassically~\cite{Brou06a}.  For these, the effect of increasing $\tE$ is twofold; first as each encounter typically lasts $\tE$, forming the diagrams considered before becomes less likely, while conversely the possibility that all the trajectories are correlated for their whole length
increases (cf the bands in \cite{Mic09}), and we need to add this additional set of diagrams.  For $n\!\leq\!3$ the results \cite{Brou06a} suggest the replacement (which is in line with effective RMT \cite{Sil03})
\beq\label{Ehrnenfestanstatz}
C(\epsilon, \tau, n) = C(\epsilon, n) \rme^{-\tau (1 -\rmi n \epsilon) } + \frac{1-\rme^{-\tau (1 -\rmi n \epsilon)}  }{1 -\rmi n \epsilon}.
\eeq
This separation into two terms was shown in \cite{Whi05} (for $\epsilon\!=\!0$) to be a consequence of the preservation under time evolution of phase-space volume inside the system and hence the splitting of phase-space into two essentially independent subsystems. 
Including an energy difference, the second term in Eq.~(\ref{Ehrnenfestanstatz}) as well as the form of the exponential in the first term follow directly.  The only thing that cannot be determined from \cite{Whi05} is the remaining factor in the first term.  To show that it is indeed $C(\epsilon,n)$ we reconsider the diagrams treated before, which were created by sliding the encounters together or into the lead, as part of a continuous deformation of a single family of diagrams.  By suitably transforming their semiclassical contributions, we can extract their complete $\tE$ dependence and show it is always $\rme^{-\tau(1-\rmi n\epsilon)}$.  Summing over all the families, this common $\tE$ dependence, along with the necessity of recovering the previous result for $\tau\!=\!0$, shows that Eq.~(\ref{Ehrnenfestanstatz}) holds for all $n$ \cite{Wal09details}.  It is possible to obtain the rest of this equation semiclassically for all $n$.  By considering correlated trajectory bands explicitly we obtain the second term, while the separation into two terms arises from opposing restrictions on the bands and the encounters \cite{Wal09details}.

Equation~(\ref{Ehrnenfestanstatz}) reproduces the two known limits: the previous RMT result
for $\tau\!=\!0$, and the BS result~\cite{Sch99,Ihr01},
$d_{\rm BS}  (\epsilon) \!=\! \left(\frac{\pi}{\epsilon}\right)^2 \frac{ \cosh\left(\pi /
\epsilon\right) }{ \sinh^2\left(\pi/\epsilon\right)}$, for $\tau \!=\! \infty$.
Alongside the two limits, this equation, and in particular the $n\epsilon$ dependence in the exponent, 
leads to interesting $\tE$-effects: a re-normalized gap, an oscillatory DoS with spikes with period $2\pi/\tau$, and a second intermediate gap.

To study this behavior, we substitute Eq.~(\ref{Ehrnenfestanstatz}) into Eq.~(\ref{densityeqn}) 
and get two contributions, from the two terms.  The first yields a reduced RMT-type contribution 
that can be evaluated as before.  The energy differential leads to two further terms, and this part of the DoS is
\beqa \label{densitysemirmt}
d_{1}(\epsilon)&=&\rme^{-\tau}\left[1-2\Real \: \rme^{\rmi\epsilon\tau}H(\epsilon,-\rme^{\rmi\epsilon\tau})\right]\\
&& {} + \tau\rme^{-\tau}\left[1-2\Real \: \rme^{\rmi\epsilon\tau}G(\epsilon,-\rme^{\rmi\epsilon\tau})\right] ,\nonumber
\eeqa
which includes the most natural constant term and reduces to (\ref{Hgeneqn}) when $\tau\!=\!0$.  Including the rest of the constant term (=$1-(1+\tau)\rme^{-\tau}$) from Eq.~(\ref{densityeqn}) with the second contribution of (\ref{Ehrnenfestanstatz}), which can be summed exactly via Poisson summation, we obtain
\begin{align} \label{densityclassic}
d_{2}(\epsilon)\!&=\!1\!-\!\left(1\!+\!\tau\right)\rme^{-\tau}\!+\!2\sum_{n=1}^{\infty}\frac{(-1)^{n}}{n}\Imag \frac{ \partial  }{\partial \epsilon}
\left[
 \frac{1\!-\!\rme^{-\tau (1\!-\!\rmi n \epsilon)}  }{1\!-\!\rmi n \epsilon} \right] \nonumber \\
&= d_{\rm BS}(\epsilon) 
-\rme^{-\frac{2\pi k}{\epsilon}}\!
 \left[ d_{\rm BS}  (\epsilon)  +
\frac{2k\, \left(\pi/\epsilon\right)^2 }{\sinh\left(\pi/\epsilon\right)}
\right], 
\end{align}
where $k=\lfloor\frac{\epsilon \tau+\pi}{2\pi} \rfloor$ involving the floor function.  We note that $d_{2}(\epsilon)$ is zero up to $\epsilon\tau\!=\!\pi$, but when combined with the first contribution, we find interesting new features.

As an illustration we plot the full DoS for $\tau \!=\! 2$ in Fig.~\ref{Figtau2}. We find a clear reduction of the RMT gap and in the inset an oscillatory behavior of the DoS at larger energy. We note that $\tE$-oscillations have previously been predicted~\cite{Ada02b,Vav03}, however those appearing here have a larger magnitude.  More interestingly though, the result in Fig.~\ref{Figtau2} shows the appearance of a second pronounced gap.  This structure in the DoS would be a clear-cut signature of the Ehrenfest time. The absence of such a feature in previous numerical work is presumably due to the difficulty in reaching the limit $\tau \geq 1$.

\begin{figure}
\centering
\includegraphics[width=0.95\columnwidth]{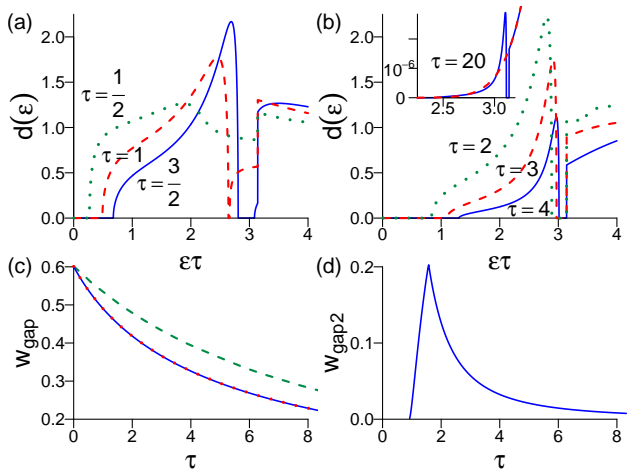}
\caption{(a), (b) Density of states as a function of $\epsilon\tau\! =\! E/E_{\rm E} $ for various values of $\tau$ 
showing the appearance of a second gap below $\epsilon\tau\! =\! \pi$. Inset: DoS for $\tau\!=\!20$
(solid line) together with the BS limit (dashed).
(c) Width of the original (first) gap as a function of $\tau$. Our semiclassical result (solid line) agrees
with effective RMT~\cite{Bee05} (dotted line); dashed line: prediction from the stochastic model 
of~\cite{Vav03}. (d) Width of the second gap as a function of $\tau$.}
\label{Figtaudiff}
\end{figure}

More generally, using Eqs.~(\ref{Gcubeqn}) and~(\ref{Hcubeqn}) for $G$ and $H$, we can express $d_{1}(\epsilon)$ in (\ref{densitysemirmt}) explicitly in a form similar to Eq.~(\ref{densityRMT}).  The result is only non-zero when 
\beqa \label{Disceqn}
D(\epsilon,\tau)&=&\epsilon^{4}-8\epsilon^{3}\sin(\epsilon\tau)+4\epsilon^{2}\left[5+6\cos(\epsilon\tau)\right] \nonumber \\
&& +24\epsilon\sin(\epsilon\tau)-8\left[1+\cos(\epsilon\tau)\right] ,
\eeqa
is positive.  $D(\epsilon,\tau)$ is negative up to the first root of Eq.~(\ref{Disceqn}), and so we see a hard gap up to this point.  As $\tau$ is increased this first gap shrinks and eventually approaches  $\epsilon\tau\!=\! \pi$, {\em i.e.}\ $E\!=\!\pi E_{\rm E}$ for $\tau\! \gg\! 1$.  We recall that the second contribution to the DoS from Eq.~(\ref{densityclassic}) is exactly zero up to this point, so if we take the limit $\tau\!\to\!\infty$ at fixed $\epsilon\tau$, we observe a hard gap up to $\pi E_{\rm E}$ in agreement with the recent complementary quasiclassical work of \cite{Mic09}. Away from this limit though, and for general $\tau$, alongside the zero contribution (up to $\epsilon\tau\!=\! \pi$) from the bands of correlated trajectories, we also have to include the contribution from the trajectories with encounters which determine the exact size of the first gap and cause the behavior described below.  We plot the width of this gap in Fig.~\ref{Figtaudiff}(c) and find exact agreement with the effective RMT prediction~\cite{Bee05,Sil03}; it also seems to be in accordance with previous numerical findings~\cite{Jac03,Kor04} limited to $\tau \!<\!1$.  

But when $\tau \! \geq \! 0.916$, the discriminant (\ref{Disceqn}) has additional roots.  In particular $D(\epsilon,\tau)$ is negative between the second and third roots so that the first contribution to the DoS again falls to zero and we see the creation of a second gap.  As $\tau$ is increased, the roots spread apart and the gap widens, but of course $d_{2}(\epsilon)$ is only zero up to $\epsilon\tau\!=\!\pi$.  For $\tau\!>\!\frac{\pi}{2}$ the third root of Eq.~(\ref{Disceqn}) is beyond this and so the second gap is cut short by the jump at $\epsilon\tau\!=\!\pi$ in the DoS coming from the second contribution in Eq.~(\ref{densityclassic}).  As $\tau$ is increased further the second gap starts to shrink, as can be seen from the plot of its width in Fig.~\ref{Figtaudiff}(d).

In Figs.~\ref{Figtaudiff}(a) and~\ref{Figtaudiff}(b) we show the DoS for different values of $\tau$, illustrating the 
formation and then the shrinking of the second gap.  Though it shrinks, the second gap persists even for large values of $\tau$ as can be seen in the inset.  Also visible in Figs.~\ref{Figtaudiff}(a) and~\ref{Figtaudiff}(b) is that the first gap slowly approaches $\epsilon\tau\! =\! \pi$: for $\tau\!=\!20$ the first hard gap ends at $\epsilon\tau\! \approx\!2.44$, but for such a large $\tau$ the contribution is so small as to be essentially indistinguishable from BS, apart from perhaps the following spike and second gap.

{\em Conclusions.}---Based on a systematic treatment of correlation functions involving $n$ scattering
matrices, we calculated the DoS of an Andreev billiard semiclassically, and recover
a hard gap extending up to $0.6E_{\mathrm T}$ as in RMT (at $\tE\!=\!0$).
Likewise, increasing $\tE$ we can see how the gap closes (approaching $E \!=\! \pi E_{\rm E}$) 
in agreement with effective RMT, and we can study the full crossover from the RMT limit to $\tE\!\gg\! \tD$.  Interestingly this transition is not smooth, and inbetween we see the formation of a second gap at  $E \!\simeq\! \pi E_{\rm E}$ for $\tE\!\simeq \!\tD$.  
Such a striking feature, which would be interesting to confirm by independent means, should be an easier $\tE$-signature to observe experimentally than the change in size of the original gap.

\begin{acknowledgments}
We thank \.{I}.~Adagideli, A.~Altland, Ph.~Jacquod, M.~Novaes, J.D.~Urbina, and R.S.~Whitney for valuable discussions.  We acknowledge funding from the DFG under GRK 638 (DW, KR), the NSF under Grant 0604859 (GB) and from the 
AvH Foundation (JK, CP).
\end{acknowledgments}

\end{document}